# Adaptive Model Refinement Approach for Bayesian Uncertainty Quantification in Turbulence Model

Fanzhi Zeng,[*] Wei Zhang,[†] Jinping Li,[†] Tianxin Zhang,[†] and Chao Yan[‡]
*National Key Laboratory of Computational Fluid Dynamics, Beihang University, Beijing, 100191, China*

The Bayesian uncertainty quantification technique has become well established in turbulence modeling over the past few years. However, it is computationally expensive to construct a globally accurate surrogate model for Bayesian inference in a high-dimensional design space, which limits uncertainty quantification for complex flow configurations. Borrowing ideas from stratified sampling and inherited sampling, an adaptive model refinement approach is proposed in this work, which concentrates on asymptotically improving the local accuracy of the surrogate model in the high-posterior-density region by adaptively appending model evaluation points. To achieve this goal, a modification of inherited Latin hypercube sampling is proposed and then integrated into the Bayesian framework. The effectiveness and efficiency of the proposed approach are demonstrated through a two-dimensional heat source inversion problem and its extension to a high-dimensional design space. Compared with the prior-based method, the adaptive model refinement approach has the ability to obtain more reliable inference results using fewer evaluation points. Finally, the approach is applied to parametric uncertainty quantification of the Menter shear–stress transport turbulence model for an axisymmetric transonic bump flow and provides convincing numerical results.

## Nomenclature

| | | |
|---|---|---|
| $c$ | = | reference length of axisymmetric transonic bump flow |
| $CD_{k\omega}$ | = | positive portion of cross-diffusion term in SST turbulence model |

---

[*] Ph. D. Candidate, School of Aeronautic Science and Engineering; zengfz_ase@buaa.edu.cn.
[†] Ph. D. Candidate, School of Aeronautic Science and Engineering.
[‡] Professor, School of Aeronautic Science and Engineering; yanchao@buaa.edu.cn (Corresponding author).

| | | |
|---|---|---|
| $C_p$ | = | pressure coefficient |
| $\mathbf{d}, \tilde{\mathbf{d}}$ | = | experimental observations and posterior predictions of quantities of interest |
| $d_n$ | = | distance from wall |
| $F_1, F_2$ | = | blending function for SST turbulence model |
| $G, G_1, G_2, G_3$ | = | source terms in numerical examples |
| $Iter$ | = | index of iteration step |
| $k$ | = | turbulent kinetic energy |
| $L_{ij}$ | = | Euclidean distance between model evaluation points |
| $M$ | = | posterior mean value |
| $Ma_\infty$ | = | Mach number of freestream |
| $N, N_p$ | = | sample size of model evaluation points and posterior samples |
| $N_d$ | = | number of QoIs |
| $p, u_j, \rho$ | = | pressure, velocity, and density |
| $P$ | = | production term of SST turbulence model |
| $r, z$ | = | radial and axial coordinates of axisymmetric transonic bump |
| $\hat{R}$ | = | Gelman–Rubin statistic criterion |
| $Re_\infty$ | = | Reynolds number of freestream |
| $s, \zeta, T$ | = | constants in numerical examples |
| $s_1, s_2, s_3, m_0, m_1, \omega_1, \omega_2, \varphi_1, \varphi_2$ | = | parameters in numerical examples |
| $S_{ij}$ | = | strain tensor |
| $t$ | = | dimensionless time |
| $\mathbf{x}, X$ | = | model evaluation point(s) |
| $x_j$ | = | Cartesian coordinates |
| $\mathbf{y}$ | = | deterministic output of computer simulation |
| $\hat{y}$ | = | estimation of computer simulation by Kriging model |
| $\alpha$ | = | relaxation factor |
| $\gamma_1, \gamma_2$ | = | derived constants in SST turbulence model |
| $\delta_{ij}$ | = | Kronecker delta function |
| $\varepsilon$ | = | Gaussian-form error term in stochastic model |
| $\theta$ | = | model parameters |
| $\kappa, a_1, \beta^*, \beta_1, \beta_2, \sigma_{\omega 1}, \sigma_{\omega 2}, \sigma_{k1}, \sigma_{k2}$ | = | closure coefficients in SST turbulence model |
| $\mu, \mu_t$ | = | molecular viscosity, and turbulent eddy viscosity |
| $\sigma, \sigma_{\text{total}}, \sigma_{\text{post}}$ | = | standard deviations of stochastic model, total error, and posterior uncertainty |
| $\Sigma$ | = | covariance matrix of stochastic model |
| $\tau_{ij}$ | = | specific Reynolds stress tensor |
| $\Phi$ | = | blend constant in SST turbulence model |
| $\chi$ | = | location of heat source |
| $\Psi_q, q$ | = | Morris–Mitchell criterion under specified $q$ |
| $\omega$ | = | dissipation per unit turbulent kinetic energy |
| $\Omega$ | = | vorticity magnitude |

## I. Introduction

IN computational fluid dynamics (CFD) simulations in aerospace engineering, Reynolds-averaged Navier–Stokes (RANS) closure models are expected to remain an essential element into the foreseeable future [1]. However, the intuitive and empirical nature of the closures in RANS turbulence modeling leads to potential limitations on accuracy and to the introduction of uncertainties, consequently reducing the credibility of the prediction [2]. It has been shown that RANS simulations do not perform particularly well in the prediction of flow configurations involving large adverse pressure gradient, separation, reattachment, etc. [3]. Duraisamy et al. [2] classified the sources of RANS model uncertainty into four levels: 1) information loss in the Reynolds-averaging process, 2) representation of the Reynolds stress as a function of the mean field, 3) the choice of the specific model functions, and 4) the coefficients of a given model. The last two levels, also called model form and parametric uncertainty respectively, are the main objects of study by turbulence modelers. Although it neglects uncertainties in the model form and is constrained by the baseline models, the parametric approach has the advantage of being non-intrusive and thus readily available to CFD practitioners [4]. In this work, we focus mainly on insufficiencies in the quantification of parametric uncertainty of RANS turbulence models.

In recent years, the development of uncertainty quantification (UQ) techniques has enabled RANS coefficients to be interpreted in probabilistic terms [5]. Statistical inference approaches based on Bayesian techniques enable data from experiments and direct numerical simulations (DNS) to be assimilated into the UQ of turbulence models, and these approaches have therefore become a focus in the field of UQ over the past few years. Cheung et al. [6] first introduced the Bayesian UQ framework to calibrate the coefficients of the Spalart–Allmaras (SA) model in an incompressible boundary layer. After that, Edeling et al. [7], Ray et al. [8], Li et al. [9], and Zhang and Fu [10, 11] conducted similar analyses for different turbulence models and flows. Further details of research on UQ in turbulence models can be found in the review by Xiao and Cinnella [4]. However, most studies to date have considered simple 2D RANS cases, and studies on 3D cases have been limited, particularly for complex geometries [12].

One of the main reasons for the limited number of applications in complex configurations is the high computational cost of CFD simulations. In the practice of aerospace engineering, it is usual that one has to run simulations for hours on a high-performance computing cluster with hundreds of CPUs. Meanwhile, Bayesian inference typically relies on posterior samplers, such as Markov chain Monte Carlo (MCMC). MCMC requires a large number of evaluations of quantities of interest (QoIs), which is an unaffordable burden for CFD simulations. Thus, computationally inexpensive approximations are used as *surrogate models* to alleviate the computational costs [13]. However, RANS turbulence models usually contain 5–10 coefficients. Model evaluation points based on priors are inevitably sparse in such a design space, and so it is still a formidable task to construct a globally sufficiently accurate surrogate model. Moreover, with the expansion in the number of evaluation points, the time expense for the construction and prediction of surrogate models increases rapidly, and therefore the Bayesian inference becomes slower.

As MCMC typically focuses on the high-posterior-density (HPD) region, which is usually a small fraction of the design space, some recent works have concentrated on local approximations of this region. Some researchers [14–16] developed multi-stage MCMC simulations, using high-fidelity polynomial chaos expansions to ensure local accuracy, while leveraging low-fidelity models to accelerate the sampling. Marzouk and colleagues [17–20] proposed MCMC samplers with local approximation. These methods use stochastic optimization or other techniques to approximately estimate the range of the HPD region, and construct a surrogate only within this area. Although these achievements are satisfactory, they have rarely been used in the context of turbulence modeling.

In contrast to previous works, the *adaptive model refinement* (AMR) approach in this article aims at improving the local accuracy of the surrogate model (*model refinement*) by adaptively appending model evaluation points to the

HPD region. The entire design space is preserved, and evaluation points are inherited. To achieve this goal, an inherited optimal Latin hypercube sampling (LHS) is developed, and it is then integrated into the Bayesian UQ framework. Through repeated iterations of model refinements and inferences, the results asymptotically approximate the "true" posterior distribution, which is inferred from the *full model* (directly solving the governing equations) rather than the surrogate. The iterations also provide a criterion for the reliability of inference results by the convergence of the posterior probability distribution. Furthermore, only the sampling strategy of evaluation points is modified, and thus the proposed approach can easily be combined with various forms of posterior sampling algorithms (e.g., the Metropolis-Hastings algorithms [21], slice sampling [22], and the no-U-turn sampler (NUTS) [23]), surrogate models (e.g., polynomial [8], Kriging [24], and neural networks [25]), and other advanced techniques (e.g., support vector machine classifiers [8] and the dimension reduction method [10]). This feature facilitates the practical use of the proposed approach.

The remainder of the paper is organized as follows. In Section II, the Bayesian UQ framework is briefly introduced, and the AMR approach for the framework is proposed. In Section III, the characteristics of the approach are demonstrated through a 2D heat source inversion problem and its high-dimensional extension. In Section IV, the proposed approach is applied to an axisymmetric transonic bump flow, which illustrates its feasibility for the Bayesian parametric inference of the Menter shear–stress transport (SST) turbulence model. Finally, conclusions are drawn in Section V.

## II. Adaptive Model Refinement Approach for Bayesian Framework

In this section, the Bayesian UQ framework is briefly described, and the theoretical outline and workflow of the AMR approach are introduced.

### A. Bayesian Uncertainty Quantification Framework

The framework of Bayesian UQ is briefly described, following previous works [6, 7]. In the Bayesian framework, uncertainties, both aleatoric and epistemic, are represented through probability. Considering existing data, the uncertainty can be quantified by the posterior probability distribution via Bayes' rule:

$$p(\boldsymbol{\theta}|\boldsymbol{d}) \propto p(\boldsymbol{d}|\boldsymbol{\theta})p(\boldsymbol{\theta}) \tag{1}$$

where $\boldsymbol{\theta}$ represents the model parameters and $\boldsymbol{d}$ is a vector containing $N_d$ scalar experimental observations. The prior $p(\boldsymbol{\theta})$ contains the prior knowledge about the parameters, and $p(\boldsymbol{d}|\boldsymbol{\theta})$ is the likelihood function. When the prior information is insufficient, a uniform distribution is usually considered as an acceptable prior, and so inferences are dominated to a great extent by the current data as the sample size increases [21]. Calculation of the likelihood function requires computational output and experimental data, which are combined by the stochastic model. Considering the characteristics of the QoIs in the test cases and the purpose of this work, namely, to verify the capability of the AMR approach, an independent and identically distributed (i.i.d) Gaussian-form error term $\varepsilon_i$ is used to model the measurement error and model error at the $i$-th observation of QoIs $d_i$. The error term $\varepsilon_i$ is defined as

$$\varepsilon_i \mid \sigma \sim N(0, \sigma^2), \qquad 1 \leq i \leq N_d \tag{2}$$

Then, $d_i$ is assumed as a combination of deterministic $y_i$ and error term $\varepsilon_i$:

$$d_i \mid \sigma, \boldsymbol{\theta} = y_i(\boldsymbol{\theta}) + \varepsilon_i, \qquad 1 \leq i \leq N_d \tag{3}$$

where the deterministic result $y_i$ is calculated by computer simulation for certain model parameters $\boldsymbol{\theta}$. The standard deviation $\sigma$ is considered as an additional parameter to be inferred. The likelihood can be written as

$$p(\boldsymbol{d} \mid \sigma, \boldsymbol{\theta}) = \frac{1}{\sqrt{(2\pi)^{N_d} |\boldsymbol{\Sigma}|}} \exp\left\{-\frac{1}{2}[\boldsymbol{d} - \boldsymbol{y}(\boldsymbol{\theta})]^T \boldsymbol{\Sigma}^{-1} [\boldsymbol{d} - \boldsymbol{y}(\boldsymbol{\theta})]\right\}$$
$$\boldsymbol{\Sigma} = \sigma^2 \boldsymbol{I} \tag{4}$$

where $|\boldsymbol{\Sigma}|$ is the determinant of $\boldsymbol{\Sigma}$.

After the stochastic model has been constructed, the posterior distribution $p(\boldsymbol{\theta}|\boldsymbol{d})$ can be obtained using MCMC sampling. The method is suitable for the estimation of high-dimensional probability distribution that cannot be expressed in analytic forms. In this work, advanced algorithms for MCMC sampling (Slice [22] and NUTS [23]) are employed to enhance the efficiency of inference. For each inference, a total of $4\times 10^4$ effective posterior samples (i.e., samples in the burn-in period are excluded) are generated from two chains in parallel. The convergence of MCMC sampling is checked via the Gelman–Rubin statistic $\hat{R}$ within and between the chains, which is suggested to be $\hat{R} < 1.1$ according to [26]. The posterior probability density is then obtained via kernel density estimation (KDE).

The effectiveness of Bayesian inference results is commonly examined by a posterior predictive check. The checking process can be carried out by propagating the posterior uncertainty of input parameters through computer simulation to obtain the posterior predictive distribution of the QoIs. The posterior probability density function (PDF) of the QoIs is expressed as

$$p(\tilde{d}\,|\,d) = \int p(\tilde{d}, \theta\,|\,d)d\theta = \int p(\tilde{d}\,|\,d, \theta)p(\theta\,|\,d)d\theta$$
$$= \int p(\tilde{d}\,|\,\theta)p(\theta\,|\,d)d\theta \tag{5}$$

where $\tilde{d}$ represents the posterior prediction of the QoIs. The integrals in Eq. (5) can be approximated using an MCMC sampler. The above algorithms are implemented via PyMC3 [27, 28].

The MCMC sampling is usually a computationally-intensive process. To construct a sufficiently accurate surrogate model for Bayesian inference in an efficient way, the AMR approach is proposed here. The theoretical outline and workflow of this proposed approach will be introduced in the rest of this section.

**B. Sampling Strategy**

Before the construction of the surrogate model, a sampling strategy that combines stratified sampling and inherited sampling is developed to generate limited model evaluation points during a sequential process. The following characteristics are considered: 1) the space-filling characteristic in high-dimensional design space; 2) the aggregation of samples in the HPD region; 3) the exploitation of existing samples.

*1. Latin hypercube sampling*

LHS is a probability-based stratified sampling strategy that stratifies the design space on the basis of marginal cumulative distribution functions (CDFs). In this work, it is appropriate to stratify marginally for the following reasons: 1) the distribution of turbulent coefficients is normal to a certain degree, and the correlation is limited (see, e.g., Refs. [6–11]); 2) the HPD region is required to be covered by model evaluation points rather than accurately captured.

The original random LHS [29] is based on the idea that only one sample in each row and each column is contained in a Latin square. In the LHS of a multivariate distribution, each variable is divided into strata of equal marginal probability $N^{-1}$, where $N$ is the sample size. In this way, the HPD regions are more densely divided. Then, a random sample is taken at each stratum. Samples from each variable are paired randomly to form a Latin square. A simple example of LHS for two variables with normal distribution from [30] is exhibited in Fig. 1 for illustration.

On the basis of random LHS, optimal LHS (OLHS) has been developed in which some features (e.g., distances [31]) are taken as optimization objectives to further improve the distribution characteristics of sampling. This idea is incorporated into the inherited optimal LHS method described below.

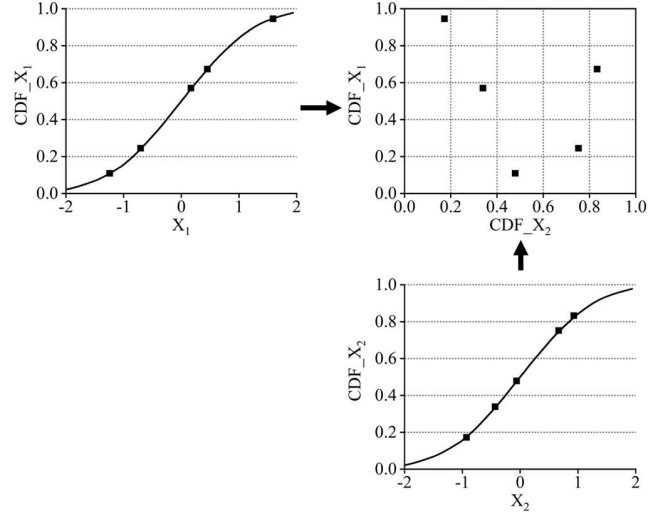

**Fig. 1** Example of LHS for two variables with normal distribution [30]

*2. Inherited optimal-LHS*

In this work, inherited optimal LHS (IOLHS) is proposed for sequential sampling in the AMR approach, which means that the AMR approach begins with fewer evaluation points and adds points sequentially during the iteration process. Inherited LHS, which was first proposed by Wang [32], uses LHS to maintain the space-filling characteristic in design space. An outline of Wang's inherited LHS is given in Algorithm 1. The computational complexity of the algorithm is $O(MK)$.

---
**Algorithm 1** Inherited random LHS
---
**Require**: $K$-dimensional design space $\Omega$; inherited sample set with $L$ evaluation points $X^I = \{x_1^I, \cdots, x_L^I\}$

1: **for** $i = 1, \ldots, K$ **do**
2:     Repartition the $i$-th dimension of $\Omega$ into a set of strata $s_i$
3:     Divide $s_i$ into two mutually exclusive subsets: strata $s_i'$ containing the elements in $X^I$, and $s_i'' = s_i \setminus s_i'$
4: **end do**
5: Calculate the maximum cardinality of $s_i''$ for $i = 1, \ldots, K$
$$M = \max_{1 \leq i \leq N} \{\text{Card}(s_i'')\}$$
6: **for** $i = 1, \ldots, K$ **do**
7:     Draw $M - \text{Card}(s_i'')$ strata from $s_i'$ randomly and add to $s_i''$
8: **end do**
9: Construct a subspace $\Omega'$ of $\Omega$ with $s_i''$ in each dimension
10: Draw $M$ samples $X^T$ in the cell center of $\Omega'$ using LHS
11: Randomly perturb the samples in $X^T$ within their cells separately, and replace to $\Omega$ as newly added samples $X^A = \{x_1^A, \cdots, x_M^A\}$
12: Combine $X^A$ and $X^I$ as $X^N$
$$X^N = \{X^A, X^I\}$$
13: **return** $X^N$
---

Let us take a 2D uniform design space as an example. The original sample set $X^I$ contains five samples, which are obtained from random-LHS (Fig. 2a). The design space $\Omega$ is divided into 5×5 equiprobability cells. To append samples, $\Omega$ is repartitioned into 8×8 equiprobability cells (Fig. 2b). The new strata that contain $X^I$ are shaded. When two inherited samples fall into the same stratum of variable $x_1$, the cardinalities of the unshaded strata in $x_1$ and $x_2$ are different, and LHS cannot be employed. Therefore, a shaded stratum is drawn to form a 4×4 subspace $\Omega'$, together with the unshaded strata. $X^T$ is sampled from $\Omega'$ (Fig. 2c). After that, $X^T$ is perturbed and then appended to the corresponding cells in $\Omega$ (blue arrows). The ultimate sample set $X^N$ is shown in Fig. 2d.

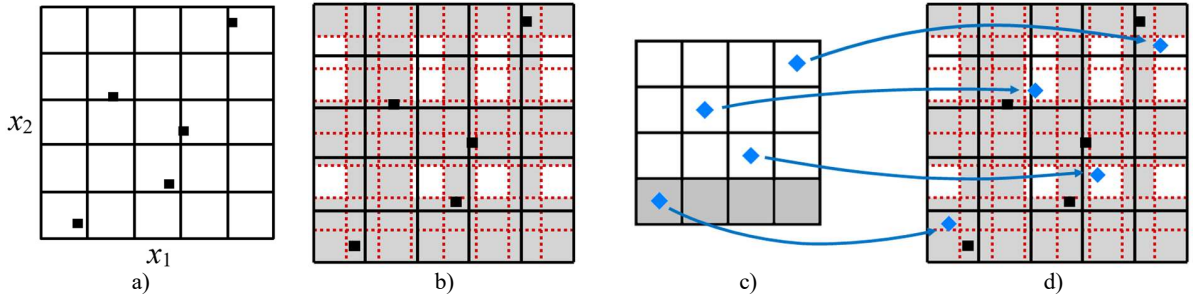

**Fig. 2 Diagram of inherited random LHS: a) Original samples $X^I$; b) Repartitioning of the design space for appending; c) Sampling of $X^T$ in subspace $\Omega'$ using LHS; d) Samples after appending**

The sampling strategy in Wang's work is only the simplest random-LHS. To further improve the uniformity of space filling and avoid the potential ill-conditioned problem of ordinary Kriging (which will be discussed in Section II.C), our strategy attempts to optimize $X^A$ to increase the distances between $X^A$ and $X^N$ by randomly changing the paring of LHS samples in $\Omega'$ (in line 10 of Algorithm 1) within finite steps. The Morris-Mitchell criterion $\Psi_q$ from [31] is used as an indicator of distance: the larger the distance, the smaller is $\Psi_q$. $\Psi_q$ is defined as

$$\Psi_q = \left( \sum_{j>i} L_{ij}^{-q} \right)^{1/q} \quad (6)$$

where $L_{ij}$ is the Euclidean distance between the $i$-th sample in $X^A$ and the $j$-th sample in $X^N$. $L_{ij}$ can be expressed as

$$L_{ij} = \left\| x_i^A - x_j^N \right\|_2 \quad (7)$$

Another problem is the choice of the specific $q$ in Eq. (6). $q$ is a positive integer. For large enough $q$, even a very small $L_{ij}$ can have a significant impact on $\Psi_q$, which is beneficial for achieving the aims of IOLHS. However, a large $q$ also leads to a lower success rate and thus increases the difficulty of optimization. Following the suggestion in [31], several searches are performed at $q$ = 1, 2, 5, 10, 20, 50, and 100, among which the result with minimum $\Psi_q$ is selected as the best design of $X^A$.

IOLHS is employed in the same example of inherited random LHS (see Fig. 3). It can clearly be seen that the optimization prevents the samples from gathering. The proposed IOLHS has been employed in a recent study [33].

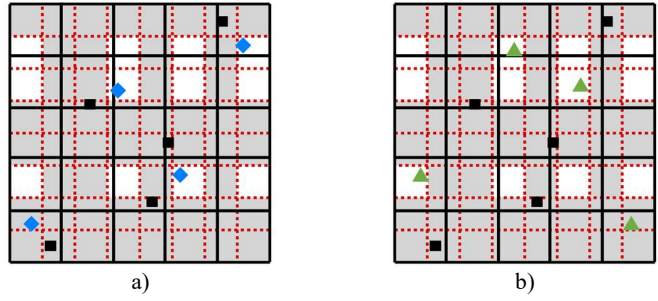

**Fig. 3 Comparison between samples from: a) inherited random LHS; b) inherited optimal LHS**

When the design space is partitioned based on the marginal CDFs, the division in regions with low probability are sparse. It is probable that the inherited samples are contained, and no sample will be added in such intervals. For that reason, the AMR approach focuses on refining the region whose posterior probability density is higher than the prior. In addition, for coefficients whose marginal posterior distribution is flat, IOLHS has the same behavior as random sampling. Therefore, the target of the AMR approach, which is referred to as the *important region* in this paper, is defined as the range of sensitive parameters with obvious probability peaks where the marginal posterior probability density is significantly higher than the prior. For a single parameter, multiple intervals may satisfy this definition. In that case, the important region only includes the intervals with probability peaks that are significantly superior to the prior distribution and other peaks.

Although the common-used definition of the HPD region [34] is different from that of the important region here, their coverages are basically consistent. Therefore, the term "HPD region," which expresses the statistical significance of this region, is still used in this paper, unless it is necessary to particularly emphasize properties related to the definition of the important region.

## C. Relaxation Factor

The sample addition process is sensitive to disturbances of the posterior distribution due to dense stratification of high-dimensional design space. Thus, too many evaluation points may be added to the temporary HPD region. Besides, evaluation points within a small distance may lead to an ill-conditioned problem in ordinary Kriging [35]. To overcome this problem, a relaxation factor $\alpha$ is introduced to smooth the posterior distribution for stratified sampling. When considering the relaxation factor $\alpha$, the stratified distribution is defined as

$$F_{SS} = (1-\alpha)F_{PPD} + \alpha F_{prior} \tag{8}$$

where $F_{SS}$, $F_{PPD}$, and $F_{prior}$ are the CDFs of the stratified distribution, the posterior and the prior, respectively. When the prior distribution is assumed to be uniform, the first derivative of the CDF of the stratified distribution is proportional to that of the posterior distribution. With increasing $\alpha$, the stratified distribution becomes flat, and the number and aggregation of newly added samples are reduced. Furthermore, tuning of the relaxation factor will not affect the location of the important region and the maximum a posteriori (MAP) estimate of the posterior.

**D. Surrogate Model**

In this work, the ordinary Kriging model is used as the surrogate model. Further details of the Kriging model can be found in [24] and [36].

For $n$ model evaluation points in $m$-dimensional design space

$$\boldsymbol{X} = \left[\boldsymbol{x}^{(1)}, ..., \boldsymbol{x}^{(n)}\right]^{\mathrm{T}} \in \mathbb{R}^{n \times m} \tag{9}$$

with corresponding responses

$$\boldsymbol{y} = \left[y^{(1)}, ..., y^{(n)}\right]^{\mathrm{T}} = \left[y(\boldsymbol{x}^{(1)}), ..., y(\boldsymbol{x}^{(n)})\right]^{\mathrm{T}} \in \mathbb{R}^{n} \tag{10}$$

the prediction of $y$ for any given $\boldsymbol{x}$ is given by

$$\hat{y}(\boldsymbol{x}) = \beta_0 + \tilde{\boldsymbol{r}}^{\mathrm{T}}(\boldsymbol{x}) \underbrace{\tilde{\boldsymbol{R}}^{-1}(\boldsymbol{y} - \beta_0 \boldsymbol{F})}_{=V_{\mathrm{krig}}} \tag{11}$$

where

$$\begin{aligned}
\boldsymbol{F} &= [1, ..., 1]^{\mathrm{T}} \in \mathbb{R}^{n} \\
\tilde{\boldsymbol{R}} &= \left(R(\boldsymbol{x}^{(i)}, \boldsymbol{x}^{(j)})\right)_{i,j} \in \mathbb{R}^{n \times n} \\
\tilde{\boldsymbol{r}} &= \left(R(\boldsymbol{x}^{(i)}, \boldsymbol{x})\right)_{i} \in \mathbb{R}^{n} \\
\beta_0 &= (\boldsymbol{F}^{\mathrm{T}} \tilde{\boldsymbol{R}}^{-1} \boldsymbol{F})^{-1} \boldsymbol{F}^{\mathrm{T}} \tilde{\boldsymbol{R}}^{-1} \boldsymbol{y}
\end{aligned} \tag{12}$$

The operator $R(\cdot, \cdot)$ is the spatial correlation function. The constant $\beta_0$ and vector $V_{\mathrm{krig}}$ in Eqs. (11) and (12) can be calculated from the model evaluation points.

**E. Workflow**

The workflow of the AMR approach can be summarized as follows.

In the beginning, OLHS is used to obtain a prior-based sample set whose size is smaller than some given value (e.g., 10 times the dimension of the design space [37]). Then, an initial surrogate model is constructed, and Bayesian inference is implemented. After the inference results have been obtained, the design space is repartitioned according to the marginal posterior probability distribution, and new evaluation points are appended using Algorithm 1. The above procedure is executed iteratively to refine the surrogate model in the current HPD region until a given convergence criterion is met. Because the accuracy of the surrogate in the low-probability-density region cannot be guaranteed, this will inevitably lead to disturbance of the posterior probability distribution during AMR. However, this disturbance should not have significant impacts on the MAP estimate and the location of the important region if the inference is convincing. Therefore, if the location of the important region and the MAP estimate remain stable, the inference results of the AMR approach can be regarded as converged. The *convergence criterion* can also serve as a criterion for the reliability of the inference results. The criterion avoids subjectivity in the selection of the error index and threshold for error analysis, which will be discussed in the context of posterior predictive check in Section III.A.2.

The workflow of the AMR approach is summarized as a flow chart in Fig. 4.

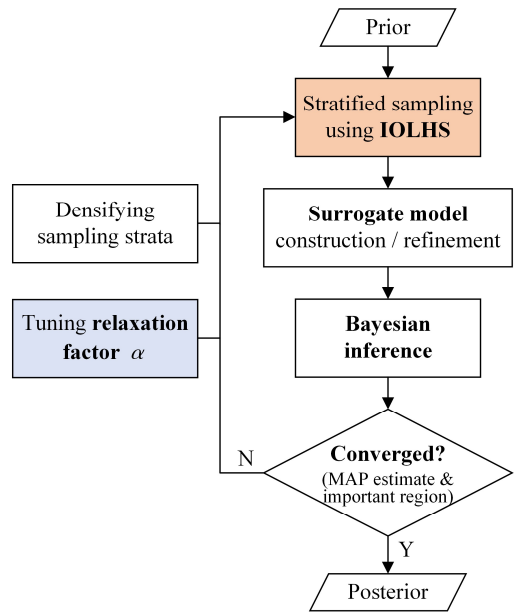

Fig. 4 Workflow of the AMR approach

In the next sections, the characteristics of the AMR approach will be demonstrated and verified through numerical examples, and the approach will then be applied to UQ of turbulence coefficients.

## III. Application in Numerical Examples

In this section, the characteristics of the AMR approach are investigated using two numerical examples: a spatially 2D heat source inversion problem from [16, 17, 38] and its extension to high-dimensional design space. In the first example, the inference is conducted in a 2D parameter space, which is beneficial for visualizing the workflow of the AMR approach. In the second example, the effectiveness and efficiency of the proposed approach are further validated. The calculations are performed using a Linux virtual machine on an Intel-i5 laptop computer.

**A. Example I: 2D Heat Source Inversion Problem**

*1. Problem Definition*

The following dimensionless diffusion equation on a spatial domain $S = [0, 1] \times [0, 1]$ with adiabatic boundaries is considered:

$$\begin{aligned}
\frac{\partial u}{\partial t} &= \nabla^2 u + G\left[1 - H(t-T)\right] \\
\nabla u \cdot \hat{\boldsymbol{n}} &= 0 \qquad \text{on } \partial S \\
u(\boldsymbol{x}, 0) &= 0
\end{aligned} \qquad (13)$$

$$G = \frac{s}{2\pi\zeta^2} \exp\left(-\frac{|\boldsymbol{\chi} - \boldsymbol{x}|^2}{2\zeta^2}\right) \qquad (14)$$

The source term $G$ in Eq. (14) is a point source with strength $s$ and width $\zeta$, located at $\boldsymbol{\chi} = (m_0, m_1)$ and active on the time interval $t \in [0, T]$. In this example, we define $s=0.5$, $\zeta=0.1$, and $T=0.05$, and leave the location $\boldsymbol{\chi}$ as an unknown parameter to be inferred. The prior distribution of $\boldsymbol{\chi}$ is assumed to be uniform in the spatial domain. The partial differential equation is solved using the finite difference method (FDM). The spatial field is discretized in a uniform grid with $h=0.025$, where the diffusion term is discretized in second-order centered differences. Time integration is performed by an explicit, third-order Runge–Kutta scheme with $\Delta t = 1 \times 10^{-4}$. The QoIs are obtained from a uniform $3 \times 3$ sensor grid at two successive times $t=0.05$ and $t=0.15$. A continuous mapping between $\boldsymbol{\chi}$ and a total of 18 QoIs can be constructed using the full model (directly solving the equations using the FDM) or surrogate models.

In this example, we set the exact value of the source location to be (0.25, 0.75). The $u$ fields at the two moments are shown in Fig. 5. The experimental data are generated by adding independent measurement noise $\varepsilon_i \sim N(0, 0.02^2)$ to the deterministic simulation result.

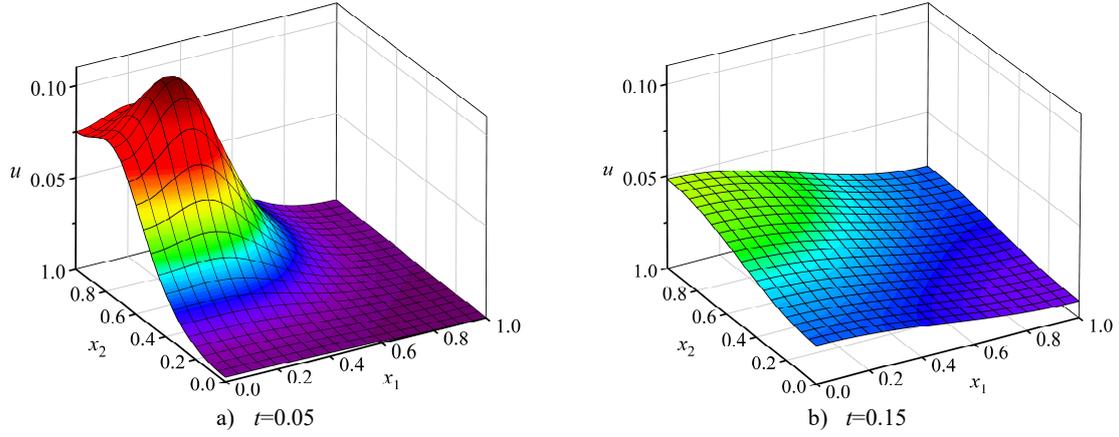

a) $t=0.05$   b) $t=0.15$

**Fig. 5 Scalar field $u$ in Example I, for $\chi = (0.25, 0.75)$**

*2. Inference Results*

In this example, the impact of the surrogate model on the Bayesian inference is first investigated. For this purpose, inference results from prior-based surrogate models are compared with those from the FDM. To construct surrogate models of different accuracies, two prior-based training sets with respective sample sizes $N=5$ and 20 are uniformly generated in design space by OLHS, where $N=20$ satisfies the recommendation in [37] and $N=5$ is seriously under-sized. Figure 6 shows a comparison among the marginal posterior distributions of parameters. Because of the random noise, the peak position of the true posterior distribution (FDM) deviates from the exact value of the source location. The MAP estimate of the standard deviation $\sigma$ in the stochastic model is approximately 0.02 (the standard deviation of the measurement noise), indicating that Bayesian inference has the ability to effectively estimate Gaussian-form uncertainty. The figure shows a huge discrepancy between the results from the lower-accuracy surrogate model ($N=5$) and FDM, and thus the result from this surrogate is unconvincing; moreover, the inference is better but is still biased when the recommended sample size is used in the construction of the higher-accuracy surrogate ($N=20$). Thus, it can be seen that the accuracy of the surrogate model is crucial to the validity of UQ results.

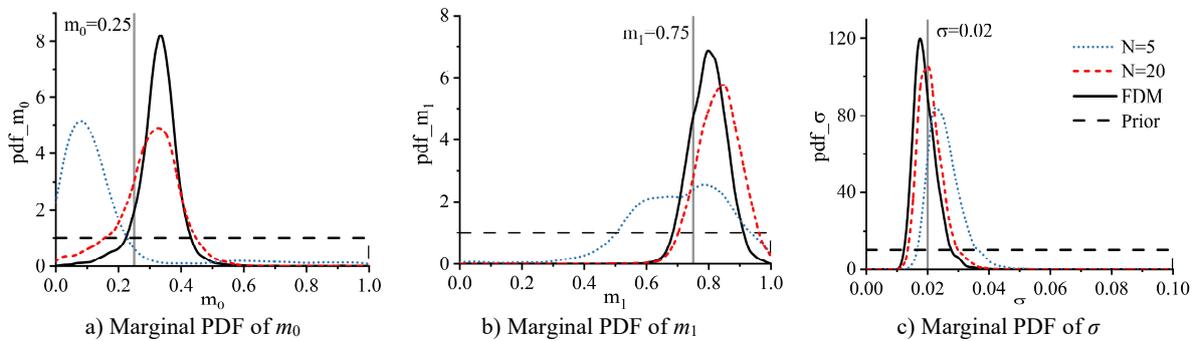

a) Marginal PDF of $m_0$   b) Marginal PDF of $m_1$   c) Marginal PDF of $\sigma$

**Fig. 6 True posterior distribution, compared with that from prior-based surrogate models for Example I**

Now, the AMR approach starts with the smaller training set $N$=5. The model refinement procedure in the first iteration step is illustrated in Fig. 7. A contour map of the joint posterior distribution is shown by the blue curves in Fig. 7a, which is obtained based on uniform stratification (green grids in Fig. 7a) and prior-based evaluation points (black dot markers). According to the marginal posterior CDFs (Fig. 7b and c), each dimension of the physical design space is repartitioned into five equiprobability strata (black grids in Fig. 7a), which form a probabilistic design space (Fig. 7d). Then, the inherited points are mapped into the probabilistic design space, and IOLHS is employed to increase the sample size. The newly added sample points (red triangle markers) are retransformed to the physical design space (Fig. 7a) by the inverse CDFs and are used to refine the surrogate model for the next iteration step. Finally, as can be seen from Fig. 7a, the model evaluation points adaptively gather in the current HPD region. After rebuilding the surrogate model, the local accuracy in this region is enhanced.

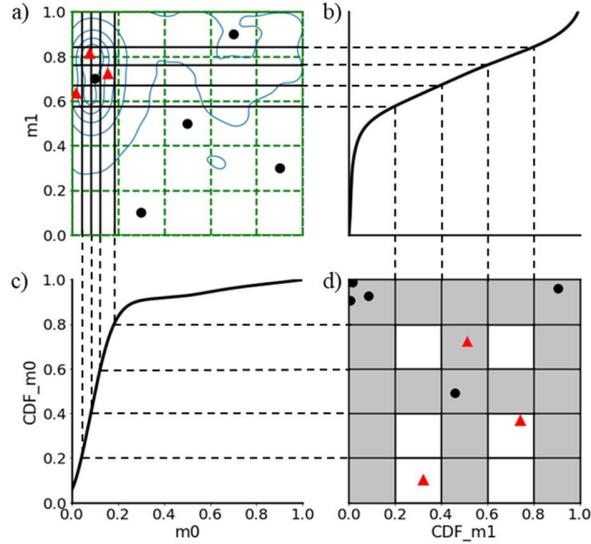

**Fig. 7 Diagram of the model refinement procedure in the first iteration step**

Following the above procedure, four iteration steps are executed until the posterior distribution converges and no samples can be added. For the purpose of confirming the convergence of the inference results, we increase the number of strata to further refine the surrogate model for the fifth step. Table 1 summarizes the total sample size, strata number and $\hat{R}$ of MCMC sampling in each iteration step (*Iter*). Figure 8 presents the locations of model evaluation points in part of the design space near the true HPD region. Notice that none of the samples in the original training sets actually falls in the HPD region. By contrast, with the AMR approach, the newly added evaluation points move toward the true HPD region and fall into this region after *Iter* 3. Figure 9 shows the inference results in each iteration step. As can be seen, with refinement of the surrogate model, the current posterior distribution gradually approximates the true distribution, and it converges after *Iter* 4. The result of *Iter* 4 shows good consistency with the true posterior distribution.

**Table 1 Summary of iteration information for Example I**

| Iteration step *Iter* | Total sample size $N$ | Strata number | $\hat{R}$ |
|---|---|---|---|
| 0 ($N$=5) | 5 | 5 | 1.02 |
| 1 | 8 | 5 | 1.04 |
| 2 | 10 | 5 | 1.02 |
| 3 | 12 | 5 | 1.01 |
| 4 | 14 | 5 | 1.01 |
| 5 | 16 | 10 | 1.01 |

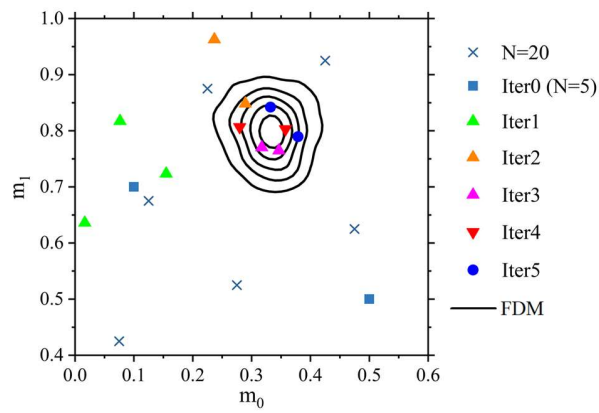

**Fig. 8 Locations of model evaluation points, together with the true joint PDF of parameters ($m_0$, $m_1$)**

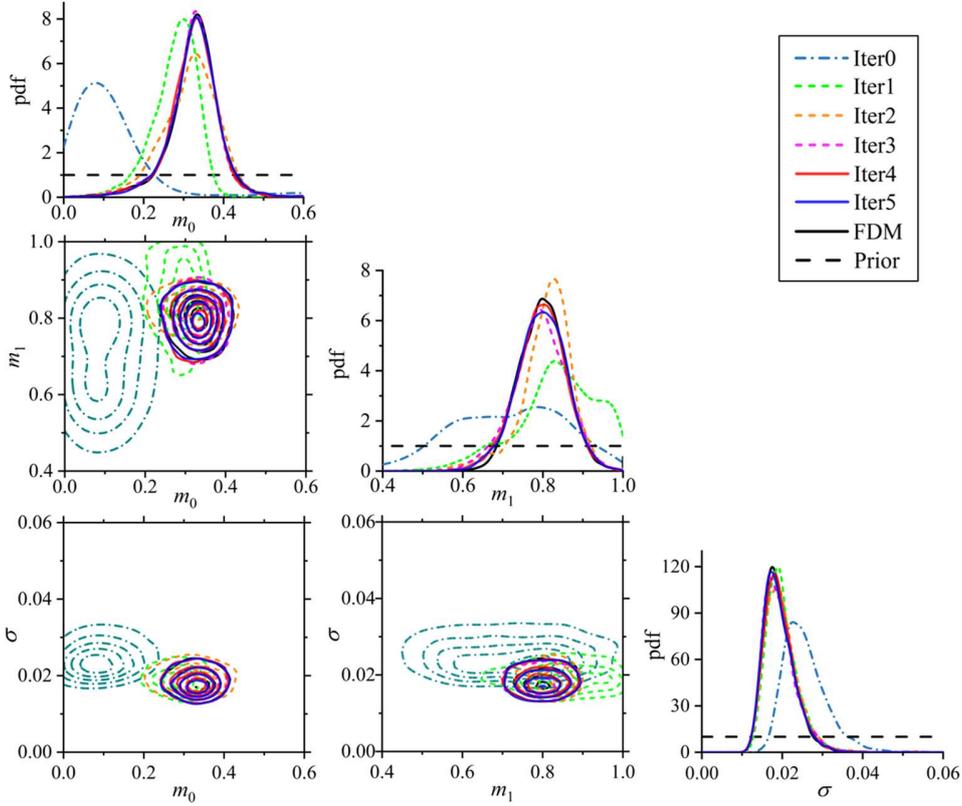

**Fig. 9** True posterior distribution, compared with that in each iteration step of the AMR approach for Example I

Finally, a posterior predictive check is executed. The surrogate errors and the discrepancies of the posterior predictive distribution between inferences using surrogates and true results are analyzed, concentrating on the QoIs at the exact source location (0.25, 0.75). The discrepancies are estimated by the Kullback–Leibler (KL) divergence [39]. The surrogate errors are estimated through the mean square error (MSE) of posterior samples, which is defined as

$$\mathrm{MSE} = \frac{1}{N_d}\frac{1}{N_p}\sum_{j=1}^{N_p}\left(y_i(\boldsymbol{\theta}_j)-S_i(\boldsymbol{\theta}_j)\right)^2 \qquad (15)$$

where $\boldsymbol{\theta}_j$ represents the $j$-th posterior samples, $S_i$ the prediction of the surrogate for the $i$-th observation, and $y_i$ the output from the FDM. $N_p$ is the size of the posterior samples, and $N_d$ is the number of QoIs. Using the MSE as the error index is appropriate here: because the values of some sensors at $t=0.05$ are very close to zero, employing the relative error would lead to numerical difficulties. However, for error analysis via the absolute error, the selection of the error threshold is case-dependent. These problems exist in all examples in this article, and this provides the motivation for the reliability criterion in the AMR approach. To obtain the MSE and the KL divergence, 5000 posterior samples are generated from each inference, respectively, using an MCMC sampler. The results are shown in Fig. 10, which uses the sample size of evaluation points *N* as the abscissa. At the beginning of the iteration, the effect of model refinement is not obvious, since the location of the newly added point is not coincident with the true HPD region. As the iteration progresses, the surrogate errors and the KL divergence are greatly reduced, and thus the reliability of the inference is greatly improved. After the iteration has converged, the MSE curve continues to decline, but the KL divergence begins to oscillate slightly within an order of magnitude. The results also show that the surrogate error of the AMR approach is lower than that of prior-based OLHS with *N*=20 (horizontal line) after *Iter* 3, which illustrates the ability to obtain more reliable inference results with fewer evaluation points than the prior-based method.

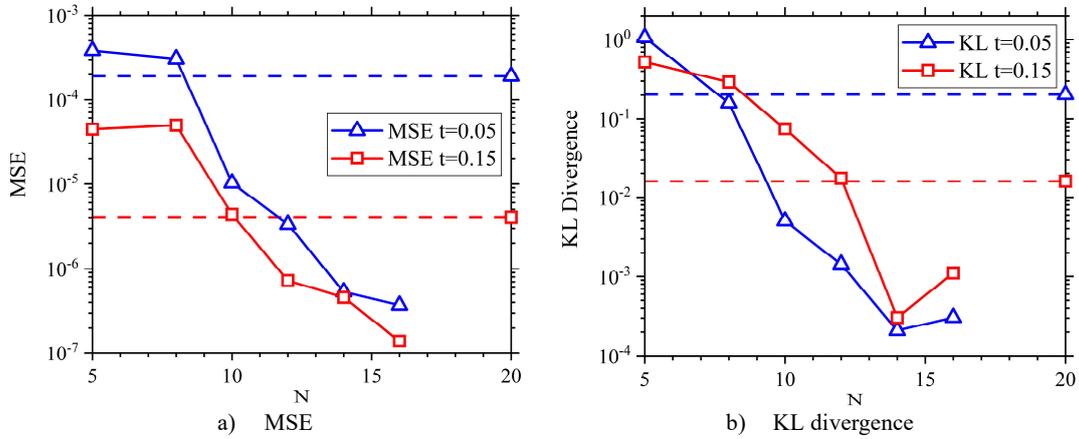

a)  MSE     b)  KL divergence

**Fig. 10 Results of posterior predictive check for Example I**

## B. Example II: Extension Example

*1. Problem Definition*

To investigate its capability and efficiency in high-dimensional design space, the AMR approach is applied to an extension of Example I in which nine parameters are to be inferred. With the governing equations being given by Eq. (13), the source term $G$ in this example is composed of a point source as in Eq. (14) together with sources in trigonometric function form in dimensions $x_1$ and $x_2$:

$$\begin{aligned}
G_1 &= \frac{s_1}{2\pi\zeta^2}\exp\left(-\frac{|\boldsymbol{\chi}-\boldsymbol{x}|^2}{2\zeta^2}\right) \\
G_2 &= s_2 \sin\left[2\pi\omega_1(x_1+\varphi_1)\right] \\
G_3 &= s_3 \sin\left[2\pi\omega_2(x_2+\varphi_2)\right] \\
G &= G_1 + G_2 + G_3
\end{aligned} \qquad (16)$$

The exact values and prior ranges of the parameters to be inferred in Eq. (16) are given in Table 2. To improve the spatial resolution for the source, the number of sensors is increased to 4×4. Only observations at $t$=0.05 are used for calibration. Other settings and the numerical scheme remain the same as in Example I.

**Table 2 Exact values and prior ranges of parameters in Example II**

| Parameter | Exact value | Lower boundary | Upper boundary |
|---|---|---|---|
| $s_1$ | 0.8 | 0.0 | 1.0 |
| $m_0$ | 0.25 | 0.0 | 1.0 |
| $m_1$ | 0.75 | 0.0 | 1.0 |
| $s_2$ | 5.0 | 0.0 | 6.0 |
| $\omega_1$ | 1.2 | 0.6 | 1.6 |
| $\varphi_1$ | 0.2 | 0.1 | 0.9 |
| $s_3$ | 4.0 | 0.0 | 6.0 |
| $\omega_2$ | 1.4 | 0.6 | 1.6 |
| $\varphi_2$ | 0.3 | 0.1 | 0.9 |

*2. Inference Results*

In this example, four prior-based training sets with respective sample sizes $N$=50, 100, 200, and 800 are used to construct surrogates with different accuracies, in which $N$=100 satisfies the recommendation [37]. The related information is summarized in Table 3. Because the HPD region is very narrow (as shown in Fig. 11), to prevent the AMR method from appending too many samples and to avoid ill-conditioned Kriging, large relaxation factors $\alpha$ are required during the whole iteration process. In this section, $\alpha$ is automatically adjusted to append about ten evaluation points according to the authors' experience.

**Table 3 Summary of iteration information in Example II**

| Type | Inference case | Total sample size $N$ | Offline time (s) | $\alpha$ | $\hat{R}$ |
|---|---|---|---|---|---|
| Prior-based | $N$=50 | 50 | 14,378 | - | 1.02 |
|  | $N$=100 | 100 | 26,944 | - | 1.02 |
|  | $N$=200 | 200 | 62,764 | - | 1.00 |
|  | $N$=800 | 800 | 300,298 | - | 1.01 |
| AMR | Iter 0 ($N$=50) | 50 | 14,378 | - | 1.02 |
|  | Iter 1 | 60 | 18,304 | 0.7 | 1.04 |
|  | Iter 2 | 68 | 19,800 | 0.5 | 1.02 |
|  | Iter 3 | 77 | 18,401 | 0.5 | 1.00 |
|  | Iter 4 | 89 | 23,177 | 0.7 | 1.01 |
|  | Iter 5 | 98 | 28,110 | 0.7 | 1.01 |
|  | Iter 6 | 109 | 32,744 | 0.7 | 1.02 |
|  | Iter 7 | 118 | 32,404 | 0.7 | 1.05 |
|  | Iter 8 | 127 | 42,053 | 0.5 | 1.02 |
|  | Iter 9 | 137 | 55,571 | 0.5 | 1.01 |

The results are presented in Fig. 11. The MAP estimates (crosses) and the important regions (error bars) of prior-based inferences (red dashed lines) and iterations during the AMR approach (blue solid lines) are given, compared with the true MAP estimates (horizontal dashed lines) and important regions (gray shading). The MAP estimates from the AMR approach are connected to show the trends. Note for some inferences that the important regions are wide

(e.g., $s_1$ in *Iter* 2), and the posterior distributions are flat. The definition of the important regions is then not strictly satisfied. For illustration, the condition is weakened to allow intervals whose probabilities are higher than the prior. It can be seen from the true results that all parameters have obvious preferences, and the HPD region occupies only a small part of the design space. Therefore, most of the prior-based evaluation points fall outside the HPD region, and even the largest sample set ($N=800$) cannot provide reliable inference results. By contrast, the AMR approach uses much fewer evaluation points to obtain results that almost completely coincide with the true results. In this example, the AMR approach converges at *Iter* 9 ($N=137$). The MAP estimates and important regions vary within 0.5% of the prior ranges in the subsequent iterations and so are not shown. Furthermore, the results basically remain unchanged in *Iter* 5 and *Iter* 6 at a position near the true results. However, an oscillation occurs at *Iter* 7, and then the results jump to the true results. This may indicate that the AMR method has a certain ability to jump out of the local optima.

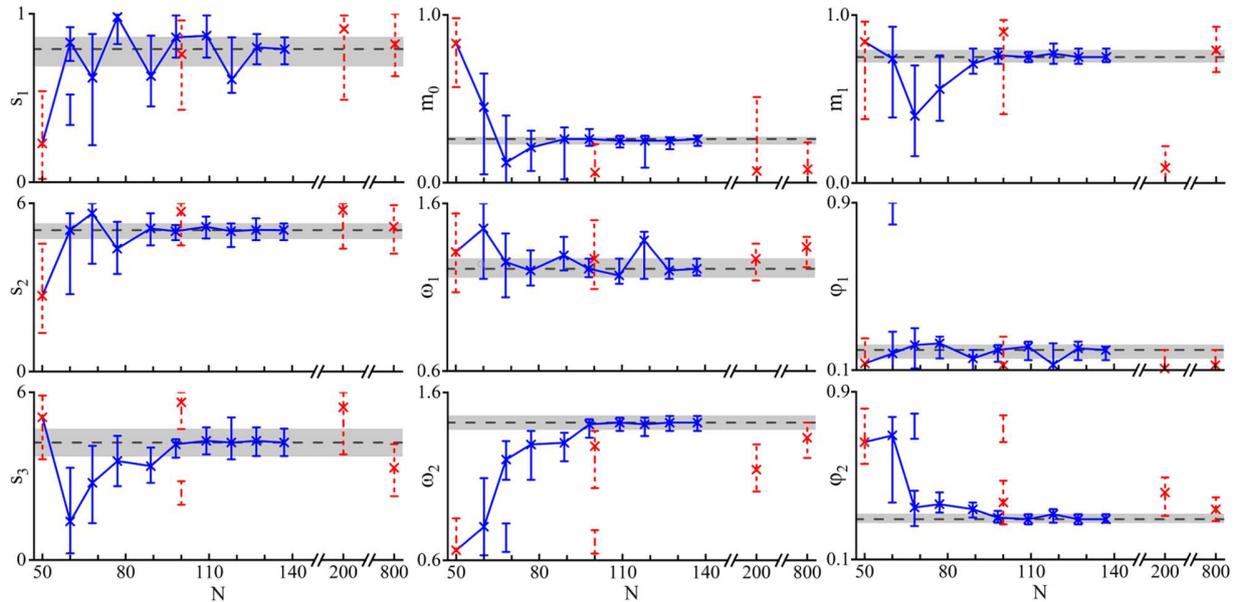

**Fig. 11 MAP estimates (crosses) and important regions (error bars) of prior-based inferences (red dashed lines) and iterations during the AMR approach (blue solid lines), compared with the true MAP estimates (horizontal dashed lines) and important region (gray shading)**

Next, the posterior predictive check is performed. In this example, we focus on examining the predictive posterior distributions of the QoIs at (0.4, 0.8), where the surrogate error is the largest. For each inference, 5000 posterior samples are exploited. Figure 12a shows the MSE curve in logarithmic coordinates. It can be seen that the surrogate error of the AMR approach decreases much more rapidly. Violin plots of the posterior predictive distributions are presented in Fig. 12b. The top and bottom ends of the violins represent the maximum and minimum values of QoIs of posterior samples. The red violins indicate that the posterior predictive distribution from the AMR approach coincides with the true result, whereas the distributions from prior-based inferences are much wider and their medians are biased.

The true posterior distributions are not always accessible owing to computation costs, and so a feasible alternative is proposed here based on the following consideration: provided that specific posterior samples are propagated through the full model and the corresponding surrogate, the probability distributions of their predictions should be similar. Posterior predictive distributions obtained by surrogate models and FDM using the same posterior samples as input are also presented in Fig. 12b (blue and red violins). Compared with prior-based inferences, the consistency of the medians and distributions from the AMR approach is significantly better. This demonstrates the practicability of the proposed approach.

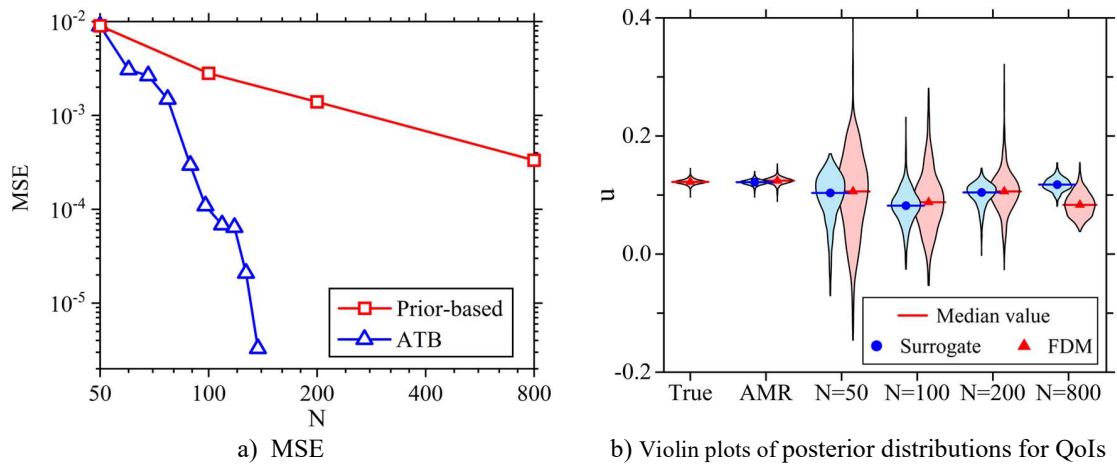

a) MSE  b) Violin plots of posterior distributions for QoIs

**Fig. 12 Results of posterior predictive check in Example II**

Finally, we discuss the efficiency of the proposed approach. The offline time costs (which include the time expenses in sampling, constructing surrogates, and MCMC sampling) are given in Table 3. A total of 284,942s is spent during the AMR approach, which is comparable to that for $N$=800, but the reliability of AMR is much better. In addition, the calculation of 663 evaluation samples is saved. Thus, the effectiveness and efficiency of the AMR approach has been demonstrated. Nevertheless, the time cost of multiple inferences must be taken into consideration. It cannot be known a priori how many iterations are needed for convergence. Therefore, it is suggested that practitioners weigh the cost between calculating evaluation points and multiple inferences before employing the AMR approach without a rigorous strategy. As far as UQ of turbulence models is concerned, the cost of inference is significantly lower than that of CFD, and thus the approach is appropriate.

# IV. Application to the SST Turbulence Model

In this section, the AMR approach is introduced into the parametric inference of the SST turbulence model for a transonic flow configuration.

## A. Menter Shear–Stress Transport Turbulence Model

The Menter shear–stress transport (SST) turbulence model [40] is a two-equation model that combines the advantage of the Wilcox $k$-$\omega$ model in the near-wall region and the robustness of the $k$-$\varepsilon$ model in regions far from the wall. The SST model is formulated as follows:

$$\frac{\partial(\rho k)}{\partial t} + \frac{\partial(\rho u_j k)}{\partial x_j} = P_1 - \beta^* \rho \omega k + \frac{\partial}{\partial x_j}\left[(\mu + \sigma_k \mu_t)\frac{\partial k}{\partial x_j}\right] \qquad (17)$$

$$\frac{\partial(\rho \omega)}{\partial t} + \frac{\partial(\rho u_j \omega)}{\partial x_j} = \frac{\gamma \rho}{\mu_t} P_1 - \beta \rho \omega^2 + \frac{\partial}{\partial x_j}\left[(\mu + \sigma_\omega \mu_t)\frac{\partial \omega}{\partial x_j}\right] + 2(1-F_1)\frac{\rho \sigma_{\omega 2}}{\omega}\frac{\partial k}{\partial x_j}\frac{\partial \omega}{\partial x_j} \qquad (18)$$

where $F_1$ is defined by

$$F_1 = \tanh\left\{\left[\min\left(\max\left(\frac{\sqrt{k}}{0.09\omega d_n}, \frac{500\mu}{\rho d_n^2 \omega}\right); \frac{4\rho \sigma_{\omega 2} k}{CD_{k\omega} d_n^2}\right)\right]^4\right\} \qquad (19)$$

with

$$CD_{k\omega} = \max\left(\frac{2\rho \sigma_{\omega 2}}{\omega}\frac{\partial k}{\partial x_j}\frac{\partial \omega}{\partial x_j}; 1\times 10^{-20}\right) \qquad (20)$$

The production term $P_1$ is approximated by

$$P_1 = \mu_t \Omega^2 \qquad (21)$$

The turbulent shear stress is modeled by the Boussinesq approximation:

$$\tau_{ij} = \mu_t\left(2S_{ij} - \frac{2}{3}\frac{\partial u_k}{\partial x_k}\delta_{ij}\right) - \frac{2}{3}\rho k \delta_{ij} \qquad (22)$$

The eddy viscosity $\mu_t$ is defined by

$$\mu_t = \min\left[\frac{\rho k}{\omega}, \frac{\rho a_1 k}{\Omega F_2}\right] \qquad (23)$$

where $F_2$ is given by

$$F_2 = \tanh\left\{\left[\max\left(\frac{2\sqrt{k}}{0.09\omega d_n}, \frac{500\mu}{\rho d^2 \omega}\right)\right]^2\right\} \tag{24}$$

The constants $\{\beta, \sigma_k, \sigma_\omega, \gamma\}$ are calculated by

$$\Phi = F_1\Phi_1 + (1-F_1)\Phi_2 \qquad (\Phi = \beta, \sigma_\omega, \sigma_k, \gamma) \tag{25}$$

where $\Phi_1$ and $\Phi_2$ represent coefficients from the Wilcox $k$-$\omega$ model and the transformed $k$-$\varepsilon$ model, respectively. $\gamma$ is calculated by

$$\begin{aligned}\gamma_1 &= \beta_1/\beta^* - \sigma_{\omega 1}\kappa^2/\sqrt{\beta^*} \\ \gamma_2 &= \beta_2/\beta^* - \sigma_{\omega 2}\kappa^2/\sqrt{\beta^*}\end{aligned} \tag{26}$$

A total of nine closure coefficients and their nominal values are given in Table 4. The prior distributions of the constants are specified as uniform distributions within about 1±40% of the nominal value. Specifically, narrower prior ranges are used for $\kappa$ and $a_1$ to reduce unphysical solutions in some parameter combinations.

**Table 4 Nominal values and prior ranges of SST coefficients**

| Coefficient | Nominal value | Lower boundary | Upper boundary |
|---|---|---|---|
| $\kappa$ | 0.41 | 0.3 | 0.5 |
| $a_1$ | 0.31 | 0.25 | 0.45 |
| $\beta^*$ | 0.09 | 0.05 | 0.14 |
| $\beta_1$ | 0.075 | 0.05 | 0.11 |
| $\sigma_{\omega 1}$ | 0.5 | 0.3 | 0.7 |
| $\sigma_{k1}$ | 0.85 | 0.5 | 1.2 |
| $\beta_2$ | 0.0828 | 0.05 | 0.12 |
| $\sigma_{\omega 2}$ | 0.856 | 0.6 | 1.2 |
| $\sigma_{k2}$ | 1.0 | 0.6 | 1.5 |

**B. Flow Configuration and Computational Details**

In this example, an axisymmetric transonic bump (ATB) flow from the NASA turbulence modeling resource (TMR) [41] is used as the flow configuration. The flow contains a shock discontinuity and separation, which are challenges for the surrogate model and the turbulence model, respectively. The geometry and computational domain are shown in Fig. 13. Only a sector is simulated, and periodic rotated boundary conditions are used on both circumferential sides of the computational domain. The Mach number $Ma_\infty$ and Reynolds number $Re_\infty$ of the freestream are 0.875 and $2\times10^6$, respectively. The experimental data on surface pressure for calibration are also provided in the TMR.

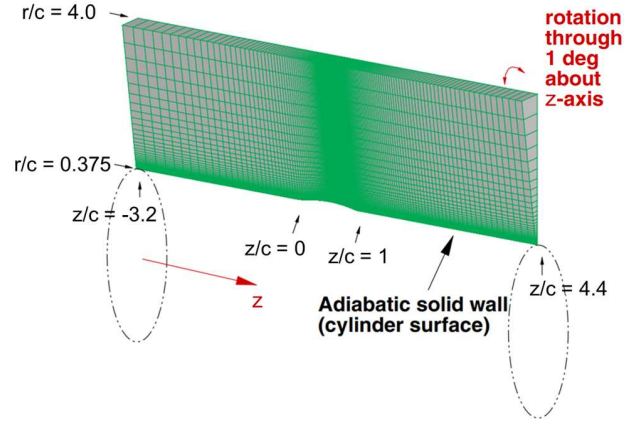

**Fig. 13 Geometry, computational domain and, mesh for ATB flow [41]**

As a compromise between accuracy and cost of CFD simulation in the process of Bayesian inference, a 2×721×321 grid from the TMR with minimum grid spacing in the wall-normal direction $y^+<1$ is used. In this work, the NASA CFL3D solver [42] is used, along with the settings given by the TMR. Figure 14 shows the surface pressure from experimental data and CFD simulation with nominal SST model coefficients, which is consistent with the results given by the TMR. The surface pressure $p$ is non-dimensionalized into pressure coefficients $C_p$, defined as

$$C_p = \frac{p - p_\infty}{0.5 \rho_\infty v_\infty^2} \qquad (27)$$

where $p_\infty$, $\rho_\infty$, and $v_\infty$ are the pressure, density, and velocity magnitude of the freestream, respectively.

Figure 15 shows the Mach number contours of the periodic side. A bow shock wave appears behind the top of the bump, which causes a very large increase in the surface pressure near $z/c=0.65$. The combination of shock and trailing-edge adverse gradient causes flow separation. The flow subsequently reattaches downstream of the bump. The nominal SST model successfully predicts the position of the pressure peak, but overestimates the pressure in the separated region.

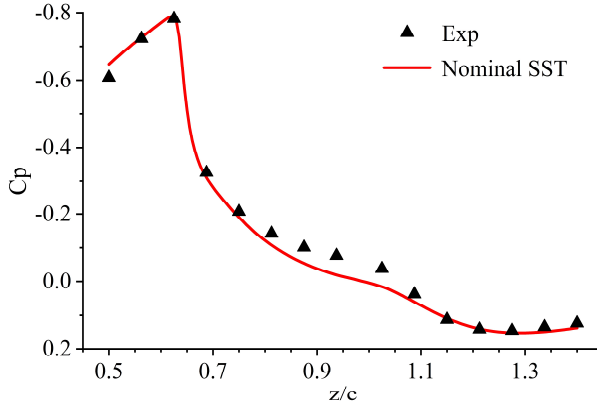 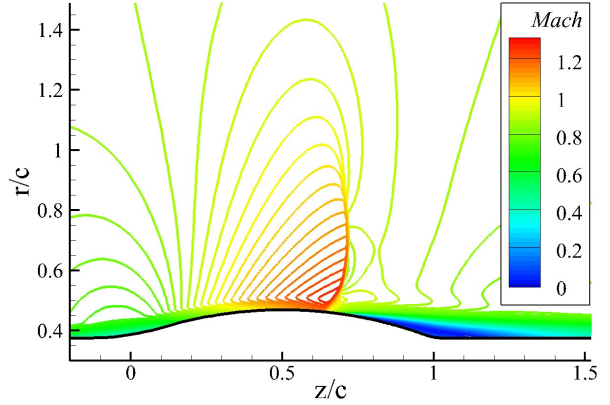

Fig. 14 Comparison of CFD simulation results of surface pressure with experimental data

Fig. 15 Mach number contours of the periodic side

### C. Prior Analysis

Before Bayesian inference is performed, a brief prior analysis is made to investigate the characteristics of $C_p$ on prior-based evaluation points and the sensitivity of surface pressure to SST model coefficients. Because the dimensionality of the design space in this section is the same as that in Section III.B, prior-based surrogate models are also generated using OLHS with training sets of sizes $N$=50, 100, and 200 ($N$=800 is excluded to reduce computational expense). The largest training set is used to ensure the accuracy of the prior analysis. The global sensitivity analysis is conducted via code modules in UQ toolkits from Sandia National Laboratories [43, 44], which uses the total Sobol index [45] as the representation of parameter sensitivity.

The $C_p$ distributions from 200 prior samples are shown in Fig. 16. As can be seen, the curves are consistent with each other in the attached flow regions. The position of the pressure peak and the surface pressure in the separated region are greatly influenced by the coefficients. As the peak moves backward, the pressure curve bulges between $z/c$=0.8 and 1.3. The correlation of the pressure distribution indicates that it might be difficult to guarantee the prediction accuracy of shock wave and separation zone at the same time by just perturbing the coefficients without changing the functional form of the SST turbulence model. The curves intersect near $z/c$=0.8, and so the uncertainty here may be relatively small. Furthermore, the Sobol indices of the coefficients for $C_p$ in the corresponding locations are presented as a stacked column chart in Fig. 16. Only the first three coefficients with the largest Sobol indices are shown individually. The chart indicates that the total Sobol indices are dominated by that of $\kappa$ and $a_1$, i.e., only $\kappa$ and $a_1$ are sensitive whereas the others are not. In this case, the notion of the important region according to the definition in Section II.B makes sense only for $\kappa$ and $a_1$. Therefore, we only present and discuss the results of these two coefficients in the subsequent study, although all nine coefficients are involved in the inference.

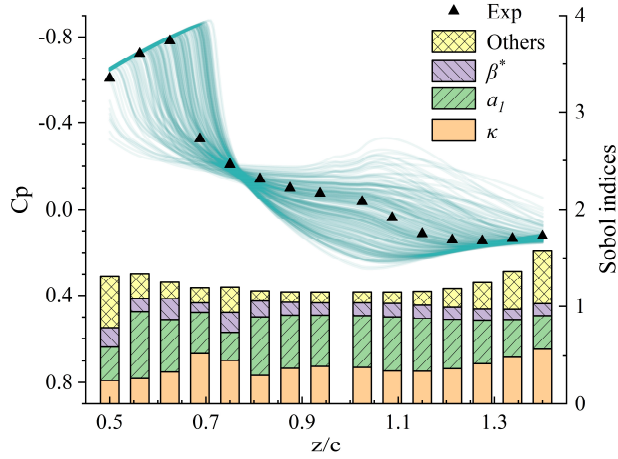

**Fig. 16 Pressure distributions and Sobol indices from 200 prior samples**

## D. Inference Results

In this case, Bayesian inferences of SST coefficients using prior-based surrogates and the AMR approach are employed. The posterior distributions from three prior-based surrogate models are shown in Fig. 17. The results for $\kappa$ (Fig. 17a) indicate that the HPD region of $\kappa$ is located on the left side of the nominal value. The results from $N$=100 and 200 agree on the probability peak of $\kappa$ at 0.326, and the result from $N$=200 shows another peak at 0.355. By contrast, the results for $a_1$ (Fig. 17b) exhibit large discrepancies. Except that the posterior distribution for $N$=200 shows a peak at 0.350, other prior-based results have no obvious preference for the value. Therefore, model refinement is needed to verify the effectiveness of the inference results.

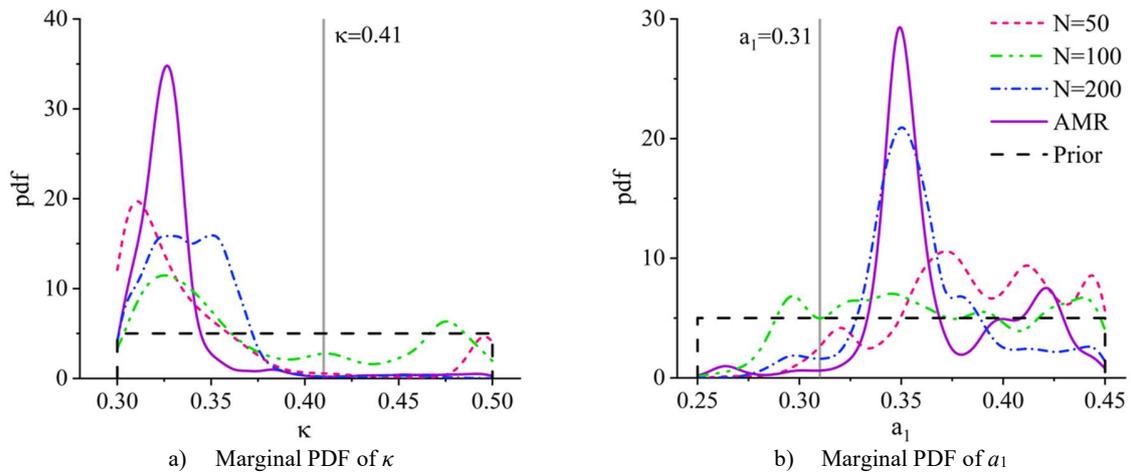

a) Marginal PDF of $\kappa$     b) Marginal PDF of $a_1$

**Fig. 17 Comparison of posterior distributions from prior-based surrogate models and the AMR approach in ATB flow**

For this problem, the AMR approach is employed on the original training set $N=50$. The related information is summarized in Table 5. The relaxation factor is also automatically adjusted to append about ten evaluation points in each iteration step. In Fig. 18, the original model evaluation points (*Iter* 0, blue triangle markers) and all appended points (red dot markers) during iterations are plotted on the $\kappa$–$a_1$ plane along with the joint PDF contour at the last iteration step. It can be seen that the original points are evenly distributed in design space, whereas the additional points are densely distributed in and around the HPD region.

**Table 5 Summary of iteration information in ATB flow**

| Iteration step *Iter* | Total sample size $N$ | $\alpha$ | $\hat{R}$ |
|---|---|---|---|
| 0 ($N$=50) | 50 | – | 1.04 |
| 1 | 59 | 0.7 | 1.02 |
| 2 | 66 | 0.7 | 1.01 |
| 3 | 74 | 0.7 | 1.02 |
| 4 | 81 | 0.5 | 1.04 |
| 5 | 89 | 0.5 | 1.02 |
| 6 | 96 | 0.3 | 1.02 |
| 7 | 103 | 0.1 | 1.01 |

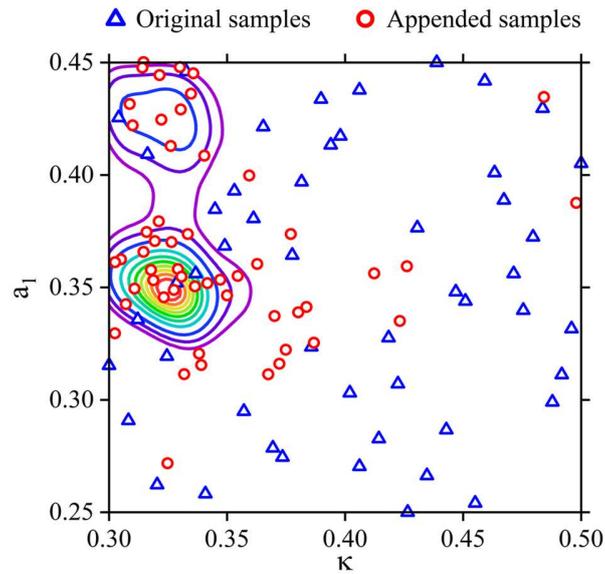

**Fig. 18 Locations of model evaluation points, together with the joint PDF obtained by the AMR approach in ATB flow**

The posterior distributions of $\kappa$ and $a_1$ at each iteration step are presented in Fig. 19. As can be seen in Fig. 19a, the MAP estimate of $\kappa$ jumps between 0.30 and 0.35 during the first four steps. Then, the MAP estimate converges to 0.326, and the important region remains stable. But the curves still fluctuate because of the disturbance in the low-probability-density region. Figure 17b shows that the MAP estimate of $a_1$ gradually approaches 0.352 and then oscillates around this position. An HPD region at high $a_1$ can be inferred from *Iter* 3 to *Iter* 5, but it is corrected by the following iteration process. The marginal PDFs of $\kappa$ and $a_1$ are given in Fig. 19c and d, which show the convergence of the posterior distribution inside the important region and the inconsistency outside this region. A comparison between the ultimate result of the AMR approach and that of prior-based surrogate models is shown in Fig. 17. It can be seen that the probability peak of $\kappa$ at 0.326 is confirmed by the AMR approach as a more credible inference result, with the other peak being abandoned. Furthermore, the HPD region of the AMR approach generally agrees with that for $N=200$, which can be seen as validation.

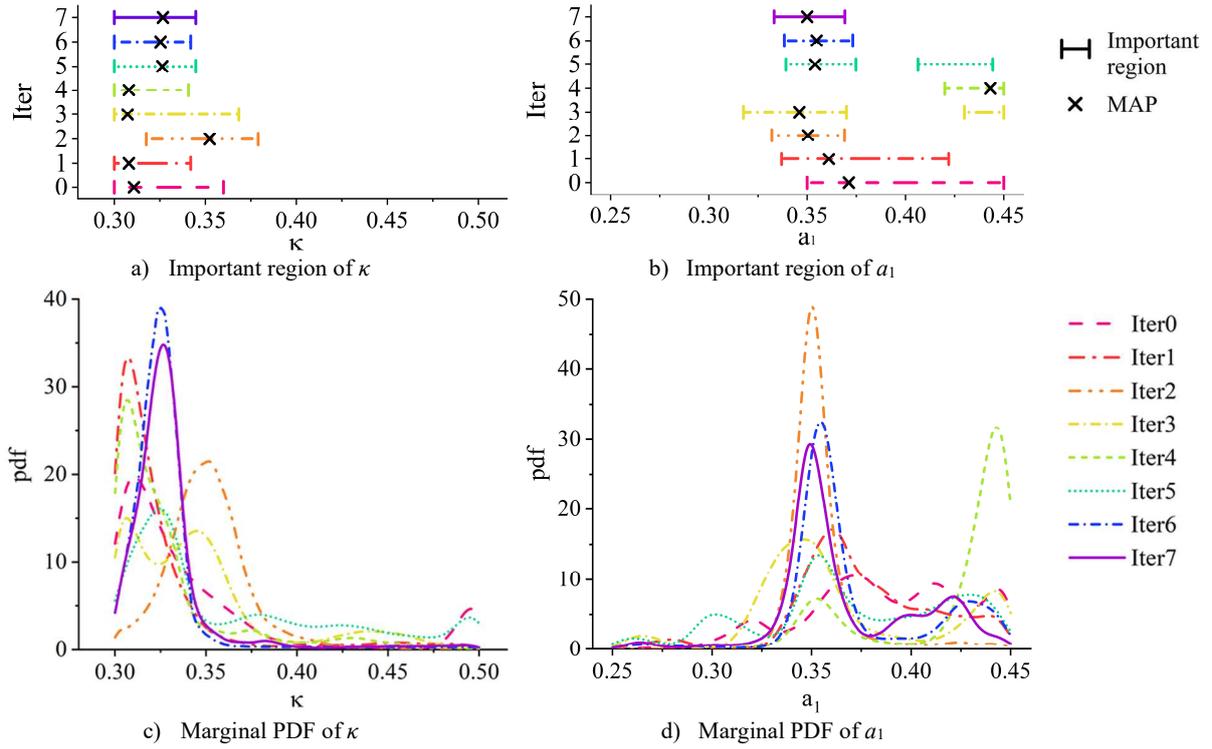

a) Important region of $\kappa$

b) Important region of $a_1$

c) Marginal PDF of $\kappa$

d) Marginal PDF of $a_1$

**Fig. 19 Posterior distributions in each iteration step of the AMR approach in ATB flow**

The posterior predictive check follows the procedure in Section III.B. In this case, 500 posterior samples are drawn from each inference. Figure 20a presents the surrogate error in each inference process, which shows a downward trend with increased number of model evaluation points. The error reduction of the AMR approach is much faster than that of the prior-based method (horizontal line), which indicates higher efficiency. Figure 20b presents violin plots and posterior medians of $C_p$ at two representative positions $z/c=0.625$ (the location of the shock wave) and $z/c=1.025$ (the maximum variance in the separated zone). The predictions in the separated zone show good consistency, whereas the surrogate error at the location of the shock wave is much higher. Compared with the results of the surrogate model, the predictive distributions obtained by CFD simulation show obvious skew and large variance. In the result for $N=50$, the predictive HPD region of CFD is much wider than that of the surrogate model, and the probability peak is indistinct. By contrast, the other two results provide accurate predictions of the MAP estimates and median values. It is noted that the improvement in this application is quite limited compared with that in Section III. One possible reason is that the parametric dimension in this case is much lower regarding the sensitivity of turbulence model coefficients, for which 200 prior-based evaluation points can almost meet the accuracy requirements. Determining the exact reason will require further study.

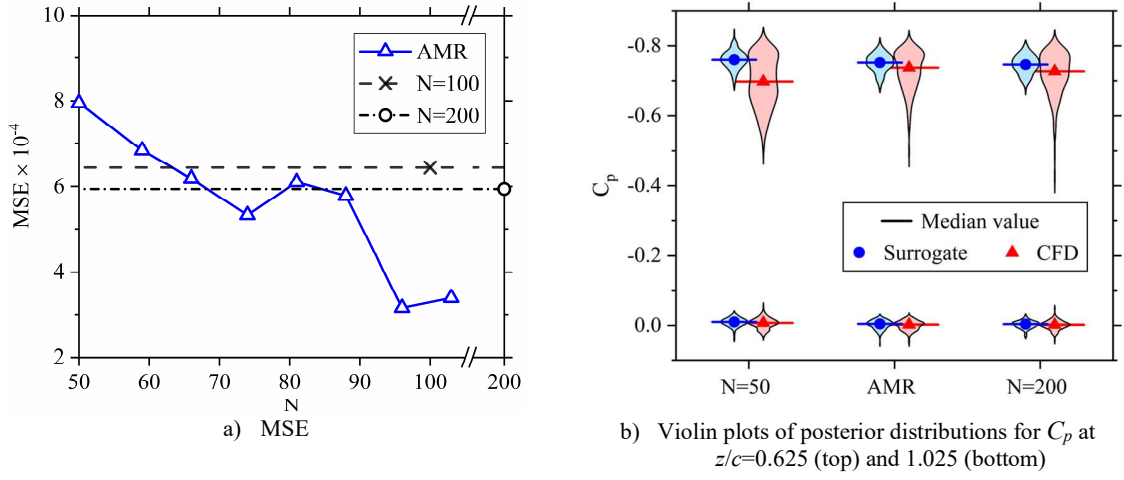

a) MSE

b) Violin plots of posterior distributions for $C_p$ at $z/c$=0.625 (top) and 1.025 (bottom)

**Fig. 20 Results of model posterior predictive check in ATB flow**

Finally, the posterior means $M$ with estimated errors calculated by CFD simulation are shown in Fig. 21. The total uncertainty of $C_p$ consists of posterior uncertainty induced by disturbance of the SST coefficients plus other errors (which are quantified by the Gaussian error term with standard deviation $\sigma$ in the stochastic model). The relationship is

$$\sigma_{\text{total}}^2 = \sigma_{\text{post}}^2 + \sigma^2 \tag{28}$$

Ranges of 1.65 times the standard deviation of the total error $\sigma_{\text{total}}$ and posterior uncertainty $\sigma_{\text{post}}$ are taken to represent the 90% credible interval. The posterior mean $M$ provides a better prediction than the nominal SST in the separated zone, but its performance is poorer at the shock wave. It can be seen as a compromise between the errors in these two regions. Moreover, the results show that the posterior uncertainties at the shock wave and in the separated zone are relatively high, which is consistent with the prior analysis. However, the credible interval of posterior uncertainty cannot cover the experimental data, and thus the errors from other sources (e.g., experimental measurement and the functional form of the SST model) should be further considered.

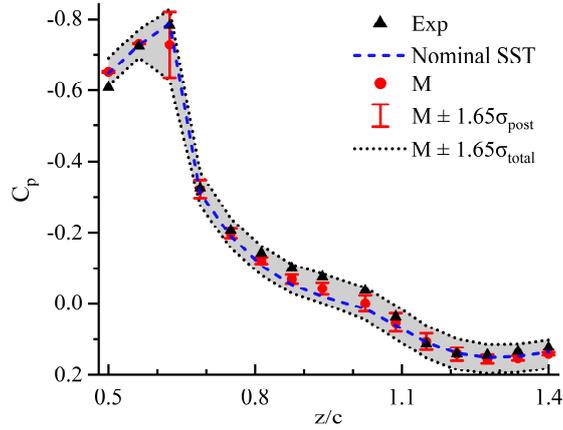

**Fig. 21 Posterior means and estimate errors for ATB flow, together with experimental data and the result of the nominal SST model**

## V. Conclusion

In this work, an adaptive model refinement (AMR) approach has been proposed to accelerate the Bayesian uncertainty quantification of turbulence model coefficients. The AMR approach has the ability to efficiently enhance the effectiveness of Bayesian inference, and provide a criterion for the reliability of inference results. The main idea is to improve the local accuracy of the surrogate model in the current HPD region by adaptively appending model evaluation points using IOLHS and thereby asymptotically approximate the true posterior distribution during iterations. To control the number of newly added evaluation points and to avoid ill-conditioned Kriging, a relaxation factor is introduced into the workflow. The reliability of the inference results can be judged from the convergence of the MAP estimates and the important regions. The approach also has the ability to jump out of local optima. These characteristics have been demonstrated in a 2D heat source inversion problem and its extension to a high-dimensional design space.

The AMR approach to turbulence models has been applied to an ATB flow involving shock discontinuity and separation, which pose challenges for the surrogate and turbulence models, respectively. The inference result shows that the surface pressure is most sensitive to $\kappa$ and $a_1$, whose MAPs are 0.326 and 0.352, respectively. The posterior mean gives a better prediction than the nominal SST model in the separated zone, but its performance is poorer at the discontinuity, which can be seen as a compromise. The 90% credible interval considering posterior uncertainty and model error overlaps with the experimental data. However, the experimental data cannot be covered only by perturbing the coefficients of the SST model, which indicates the necessity for further analysis of uncertainties arising from other sources.

Although the effectiveness and efficiency of the proposed approach are convincing, the time cost of multiple inferences is considerable. Therefore, the cost of calculating evaluation points and multiple inferences should be weighed carefully before employing the AMR approach. The current study has also found that the stratified sampling is over-sensitive to disturbances of the posterior probability distribution due to dense stratification in the high-dimensional design space. Therefore, a practical way to further improve robustness and efficiency might be to couple the AMR approach with a dimension-reduction method. Further work will also involve more applications to computationally intensive complex flows in aerospace engineering.

## Acknowledgments

This research work is supported by the National Numerical Wind Tunnel Project of China (No. NNW2019ZT1-A03), and the National Natural Science Foundation of China (No. 11721202).

## References

[1] Durbin, P. A., "Some Recent Developments in Turbulence Closure Modeling," *Annual Review of Fluid Mechanics*, Vol. 50, 2018, pp. 77–103.
https://doi.org/10.1146/annurev-fluid-122316-045020

[2] Duraisamy, K., Iaccarino, G., and Xiao, H., "Turbulence Modeling in the Age of Data," *Annual Review of Fluid Mechanics*, Vol. 51, 2019, pp. 357–377.
https://doi.org/10.1146/annurev-fluid-010518-040547

[3] Bush, R. H., Chyczewski, T., Duraisamy, K., Eisfield, B., Rumsey, C. L., and Smith, B. R., "Recommendations for Future Efforts in RANS Modeling and Simulation," AIAA Paper 2019–0317, Jan. 2019.
https://doi.org/10.2514/6.2019-0317

[4] Xiao, H., and Cinnella, P., "Quantification of model uncertainty in RANS simulations: A review," *Progress in Aerospace Sciences*, Vol. 108, July 2019, pp. 1–31.
https://doi.org/10.1016/j.paerosci.2018.10.001

[5] Duraisamy, K., Spalart, P. R., and Rumsey, C. L., "Status, Emerging Ideas and Future Directions of Turbulence Modeling Research in Aeronautics," NASA/TM–2017–219682, 2017.

[6] Cheung, S. H., Oliver, T. A., Prudencio, E. E., Prudhomme, S., and Moser, R. D., "Bayesian uncertainty analysis with applications to turbulence modeling," *Reliability Engineering and System Safety*, Vol. 96, No.9, 2011, pp. 1137–1149.
https://doi.org/10.1016/j.ress.2010.09.013

[7] Edeling W N, Cinnella P, Dwight R P, and Bijl, H., "Bayesian estimates of parameter variability in the $k$-$\varepsilon$ turbulence model," *Journal of Computational Physics*, Vol. 258, Feb. 2014, pp. 73–94.
https://doi.org/10.1016/j.jcp.2013.10.027

[8] Ray, J., Lefantzi, S., Arunajatesan, S., and Dechant, L., "Bayesian parameter estimation of a $k$-$\varepsilon$ model for accurate Jet-in Crossflow Simulations," *AIAA Journal*, Vol. 54, No. 8, 2016, pp. 2432–2448.
https://doi.org/10.2514/1.J054758

[9] Li, J., Zeng, F., Chen, S., Zhang, K., and Yan, C., "Bayesian model evaluation of three $k$–$\omega$ turbulence models for hypersonic shock wave–boundary layer interaction flows," *Acta Astronautica*, Vol. 189, 2021, pp.143–157.
https://doi.org/10.1016/j.actaastro.2021.08.050

[10] Zhang, J., and Fu, S., "An efficient Bayesian uncertainty quantification approach with application to $k$-$\omega$-$\gamma$ transition modeling," *Computers and Fluids*, Vol. 161, Jan. 2018, pp. 211–224.
https://doi.org/10.1016/j.compfluid.2017.11.007

[11] Zhang, J., and Fu, S., "An efficient approach for quantifying parameter uncertainty in the SST turbulence model," *Computers and Fluids*, Vol. 181, Mar. 2019, pp. 173–187.
https://doi.org/10.1016/j.compfluid.2019.01.017

[12] Slotnick, J., Khodadoust, A., Alonso, J., Darmofal, D., Gropp, W., Lurie, E., and Mavriplis, D., "CFD Vision 2030 Study: A Path to Revolutionary Computational Aerosciences," NASA/CR–2014–218178, 2014.

[13] Kennedy, M. C., and O'Hagan, A., "Bayesian calibration of computer models," *Journal of the Royal Statistical Society: Series B (Statistical Methodology)*, Vol. 63, No. 3, 2001, pp. 425–464.
http://doi.org/10.1111/1467-9868.00294

[14] Christen, J. A., and Fox, C., "Markov chain Monte Carlo using an approximation," *Journal of Computational and Graphical Statistics*, Vol. 14, No. 4, 2005, pp. 795–810.
http://doi.org/10.1198/106186005X76983

[15] Efendiev, Y., Hou, T., and Luo, W., "Preconditioning Markov chain Monte Carlo simulation using coarse-scale models," *SIAM Journal on Scientific Computing*, Vol. 28, No. 2, 2006, pp. 776–803.
https://doi.org/10.1137/050628568

[16] Liang, Y., and Tao, Z., "Adaptive multi-fidelity polynomial chaos approach to Bayesian inference in inverse problems," *Journal of Computational Physics*, Vol. 381, No. 15, 2019, pp. 110–128.
http://doi.org/10.1016/j.jcp.2018.12.025

[17] Li, J., and Marzouk, Y. M., "Adaptive construction of surrogates for the Bayesian solution of inverse problems," *SIAM Journal on Scientific Computing*, Vol. 36, No. 3, 2014, pp. A1163–A1186.
http://doi.org/10.1137/130938189

[18] Conrad, P. R., Marzouk, Y. M., Pillai, N. S., and Smith, A., "Accelerating asymptotically exact MCMC for computationally intensive models via local approximations", *Journal of the American Statistical Association*, Vol. 111, No. 516, 2016, pp. 1591–1607.
http://doi.org/10.1080/01621459.2015.1096787

[19] Cui, T., Marzouk, Y. M., and Wilcox, K., "Data-driven model reduction for the Bayesian solution of inverse problems", *International Journal for Numerical Methods in Engineering*, Vol. 102, No. 5, 2015, pp. 966–990.
http://doi.org/10.1002/nme.4748

[20] Conrad, P. R., David, P. R., Marzouk, Y. M., Pillai, N. S., and Smith, A., "Parallel local approximation MCMC for expensive models," *SIAM/ASA Journal on Uncertainty Quantification*, Vol. 6, No. 1, 2018, pp. 339–373.
http://doi.org/10.1137/16M1084080

[21] Gelman, A., Carlin, J. B., Stern, H. S., Dunson, D. B., Vehtari, A., and Rubin, D. B., *Bayesian data analysis*, 3rd ed., Chapman and Hall/CRC Press, New York, 2013, Chaps. 11, 2.
https://doi.org/10.1201/b16018

[22] Neal, R. M., "Slice sampling," *Annals of Statistics*, Vol. 31, No. 3, 2003, pp. 705–767.
http://doi.org/10.1214/aos/1056562461

[23] Hoffman, M. D., and Gelman, A., "The no-U-turn sampler: Adaptively setting path lengths in Hamiltonian Monte Carlo," *Journal of Machine Learning Research*, Vol. 15, No. 1, 2014, pp. 1593–1623.
https://arxiv.org/abs/1111.4246

[24] Krige, D. G., "A Statistical Approach to Some Basic Mine Valuation Problems on the Witwatersrand," *Journal of the Chemical Metallurgical & Mining Society of South Africa*, Vol. 94, No. 3, 1951, pp. 95–111.
http://doi.org/10520/AJA0038223X_4858

[25] He, X., Zhao, F., and Vahdati, M., "Uncertainty Quantification of Spalart–Allmaras Turbulence Model Coefficients for Simplified Compressor Flow Features," *Journal of Fluids Engineering*, Vol. 142, No. 9, 2019, pp. 091501.
https://doi.org/10.1115/1.4047026

[26] Gelman, A., and Rubin, D. B., "Inference from iterative simulation using multiple sequences," *Statistical Science*, Vol. 7, No. 4, 1992, pp. 457–472.

http://doi.org/10.1214/ss/1177011136

[27] Salvatier, J., PyMC3, Ver. 3.10.0, 2020.

https://github.com/pymc-devs/pymc3/releases/tag/v3.10.0

[28] Salvatier, J., Wiecki, T. V., and Fonnesbeck, C., "Probabilistic programming in Python using PyMC3," *PeerJ Computer Science*, 2:e55, 2016.

http://doi.org/10.7717/peerj-cs.55

[29] McKay, M. D., Beckman, R. J., and Conover, W. J., "Comparison of Three Methods for Selecting Values of Input Variables in the Analysis of Output from a Computer Code," *Technometrics*, Vol. 21, No. 2, 1979, pp. 239–245.

http://doi.org/10.1080/00401706.1979.10489755

[30] Minasny, B., and McBratney, A. B., "A conditioned Latin hypercube method for sampling in the presence of ancillary information," *Computers & Geosciences*, Vol. 32, No. 9, 2006, pp. 1378–1388.

https://doi.org/10.1016/j.cageo.2005.12.009

[31] Morris, M. D., and Mitchell, T. J., "Exploratory designs for computational experiments," *Journal of Statistical Planning and Inference*, Vol. 43, No. 3, 1995, pp. 381–402.

https://doi.org/10.1016/0378-3758(94)00035-T

[32] Wang, G. G., "Adaptive Response Surface Method Using Inherited Latin Hypercube Design Points," *Journal of Mechanical Design*, Vol. 125, No. 2, 2003, pp. 210–220.

https://doi.org/10.1115/1.1561044

[33] Zhang, W., Wang, Q., Zeng, F., and Yan, C., "An adaptive Sequential Enhanced PCE approach and its application in aerodynamic uncertainty quantification," *Aerospace Science and Technology*, Vol. 117, Oct. 2021, 106911.

https://doi.org/10.1016/j.ast.2021.106911

[34] Edeling, W. N., Cinnella, P., and Dwight, R. P., "Predictive RANS simulations via Bayesian Model-Scenario Averaging," *Journal of Computational Physics*, Vol. 275, Oct. 2014, pp. 65–91.

https://doi.org/10.1016/j.jcp.2014.06.052

[35] Abadou, R., Bagtzoglou, A. C., and Wood, E. F., "On the Condition Number of Covariance Matrices in Kriging, Estimation, and Simulation of Random Fields," *Mathematical Geology*, Vol. 26, No.1, 1994, pp. 99–133.

https://doi.org/10.1007/BF02065878

[36] Sacks, J., Welch, W. J., Mitchell, T. J., and Wynn, H. P., "Design and Analysis of Computer Experiments," *Statistical Science*, Vol. 4, No. 4, 1989, pp. 409–423.

https://doi.org/10.1214/ss/1177012413

[37] Loeppky, J. L., Sacks, J., and Welch, W. J., "Choosing the Sample Size of a Computer Experiment: A Practical Guide," *Technometrics*, Vol. 51, No. 4, 2009, pp. 366–376.

http://doi.org/10.1198/TECH.2009.08040

[38] Marzouk, Y. M., Najm, H. N., and Rahn, L. A., "Stochastic spectral methods for efficient Bayesian solution of inverse problems," *Journal of Computational Physics*, Vol. 224, No. 2, 2007, pp. 560–586.

[39] Kullback, S., and Leibler, R. A., "On Information and Sufficiency," *The Annals of Mathematical Statistics*, Vol. 22, No. 1, 1951, pp. 79–86.

https://doi.org/10.1214/aoms/1177729694

[40] Menter, F. R., "Two-equation eddy-viscosity turbulence models for engineering applications," *AIAA Journal*, Vol. 32, No. 8, 1994, pp. 1598–1605.

http://doi.org/10.2514/3.12149

[41] Rumsey, C. L., Turbulence Modeling resource, NASA Langley Research Center, Hampton, VA, 2020.

https://turbmodels.larc.nasa.gov/axibump_val.html

[42] Rumsey, C. L., CFL3D, Ver. 6.7, NASA Langley Research Center, Hampton, VA, 2020.

https://cfl3d.larc.nasa.gov/

[43] Debusschere, B. J., UQ Toolkit, Ver. 3.1.0, Sandia National Laboratories, Albuquerque, NM, 2020.

https://www.sandia.gov/UQToolkit/index.html

[44] Debusschere, B. J., Najm, H. N., Pebay, P. P., Knio, O. M., Ghanem, R. G., and Maitre, O. P., "Numerical Challenges in the Use of Polynomial Chaos Representations for Stochastic Processes," *SIAM Journal on Scientific Computing*, Vol. 26, No. 2, 2005, pp. 698–719.

https://doi.org/10.1137/S1064827503427741

[45] Sudret, B., "Global sensitivity analysis using polynomial chaos expansions," *Reliability Engineering and System Safety*, Vol. 93, No. 7, 2008, pp. 964–979.

https://doi.org/10.1016/j.ress.2007.04.002